\documentclass[sigplan,10pt]{acmart}
\renewcommand\footnotetextcopyrightpermission[1]{}

\startPage{1}

\setcopyright{none}

\usepackage{booktabs}   
\usepackage{subcaption} 
\usepackage{amsmath}
\DeclareMathOperator{\where}{\textbf{where}}
\newcommand{\pluseq}{\mathrel{+}=}

\newcommand{\nnz}{\text{nnz}}
\newcommand{\fhash}{\mathbf{h}}
\newcommand{\flist}{\mathbf{c}}
\newcommand{\farray}{\mathbf{u}}
\newcommand{\pstep}{\mathbf{s}}
\newcommand{\plocate}{\mathbf{l}}
\newcommand{\pinsert}{\mathbf{i}}
\newcommand{\pappend}{\mathbf{a}}
\usepackage{algpseudocode}
\usepackage{tikz}
\usetikzlibrary{calc, positioning, shapes, arrows}
\algtext*{EndWhile}
\algtext*{EndFor}
\algtext*{EndIf}
\usepackage{makecell}
\usepackage{multicol}

\iffalse
\newcommand{\TODO}[1]{}
\newcommand{\saman}[1]{}
\newcommand{\peter}[1]{}
\newcommand{\fred}[1]{}
\else
\newcommand{\TODO}[1]{{\color{red}#1}}
\newcommand{\saman}[1]{{\color{purple}Saman: #1}}
\newcommand{\peter}[1]{{\color{brown}Peter: #1}}
\newcommand{\fred}[1]{{\color{blue}Fred: #1}}
\fi

\begin{document}

\title[An Asymptotic Cost Model for Autoscheduling Sparse Tensor Programs]{An Asymptotic Cost Model for \\ Autoscheduling Sparse Tensor Programs}

\author{Peter Ahrens}
\affiliation{
  \department{Computer Science and Artificial Intelligence Laboratory}              
  \institution{Massachusetts Institute of Technology}            
}
\email{pahrens@csail.mit.edu}          

\author{Fredrik Kjolstad}
\affiliation{
  \department{Department of Computer Science}             
  \institution{Stanford University}           
}
\email{kjolstad@cs.stanford.edu}         

\author{Saman Amarasinghe}
\affiliation{
  \department{Computer Science and Artificial Intelligence Laboratory}              
  \institution{Massachusetts Institute of Technology}            
}
\email{saman@csail.mit.edu}

\begin{abstract}
While loop reordering and fusion can make big impacts on the constant-factor
performance of dense tensor programs, the effects on sparse tensor programs are
asymptotic, often leading to orders of magnitude performance differences in
practice. Sparse tensors also introduce a choice of compressed storage formats
that can have asymptotic effects.  Research into sparse tensor compilers
has led to simplified languages that express these tradeoffs, but the user is
expected to provide a schedule that makes the decisions.  This is challenging
because schedulers must anticipate the interaction between sparse formats, loop
structure, potential sparsity patterns, and the compiler itself.  Automating
this decision making process stands to finally make sparse tensor compilers
accessible to end users.

We present, to the best of our knowledge, the first automatic asymptotic
scheduler for sparse tensor programs. We provide an approach to abstractly
represent the asymptotic cost of schedules and to choose between them.  We
narrow down the search space to a manageably small ``Pareto frontier'' of asymptotically
undominated kernels. We test our approach by compiling these kernels with the TACO sparse tensor compiler and comparing them with those generated with the default TACO schedules. Our results show that our approach reduces the scheduling space by orders of magnitude and that the generated kernels perform asymptotically better than those generated using the default schedules.
\end{abstract}

\ccsdesc[500]{Software and its engineering~General programming languages}
\ccsdesc[300]{Social and professional topics~History of programming languages}

\keywords{Sparse Tensors, Compilers, Automatic Scheduling, Conjunctive Query Containment, Query Optimization}

\maketitle
\pagestyle{plain}

\section{Introduction}

Transformations like loop reordering or loop fusion can
have large effects on constant factors in the runtime of dense tensor
programs \cite{anderson_efficient_2021}.  However, the same transformations can
have even larger asymptotic effects when tensors are sparse.

As we illustrate in Section \ref{sec:motivatingexample}, loop reordering constitutes the main asymptotic difference between the three main algorithms for sparse-sparse matrix multiply (SpGEMM) \cite{mcnamee_algorithm_1971, gustavson_two_1978, buluc_representation_2008}.
 To see the effects of
loop fusion, consider the sampled matrix multiply $A_{ij} = \sum_k B_{ik} \cdot
C_{kj} \cdot D_{ij}$ where $B$ and $C$ are dense and $A$ and $D$ are sparse
(SDDMM). When we multiply $B$ and $C$ first, then multiply by $D$, our runtime
is $O(IJK)$, where $I$, $J$, and $K$ are the dimensions of their lowercase
variables.  When we multiply all three kernels in the same nested loop, our
runtime is $O(\nnz(D)K)$, where $\nnz(D)$ is the number of nonzeros of $D$, much
less than $IJ$. Complicating matters, sparse tensors may be stored in different
compressed formats with varied asymptotic behaviors.  Making $n$ random updates
to a list of nonzero coordinates might take $O(n^2)$ time, but would only take
$O(n)$ time if the coordinates were stored in a hash table.

Research into sparse tensor compilers has led to simplified languages to express
these tradeoffs and generate efficient implementations
\cite{kotlyar_compiling_1997, kotlyar_relational_1997-1,
kotlyar_relational_1999, bik_automatic_1994,bik_compilation_1993,
pugh_sipr_1999, strout_approach_2016, arnold_data-parallel_2011,
venkat_loop_2015, kjolstad_tensor_2017, kjolstad_tensor_2019,
chou_automatic_2020, chou_format_2018,
henry_compilation_2021,senanayake_sparse_2020, tian_high-performance_2021}.
These sparse tensor compilers separate \textbf{mechanism} (how code is generated
from the high-level description) from \textbf{policy} (deciding what high-level
description is best) \cite{ragan-kelley_halide_2013}. 

Sparse tensor compilers leave the policy to the user, which is often too great a
burden.  Writing a good schedule requires time and expertise. The user might
need to schedule too many kernels, or may be unfamiliar with the intricacies of
the tensor compiler in question. Sparse systems must follow the lead of dense systems, which are now moving towards automatic
scheduling \cite{anderson_efficient_2021,adams_learning_2019,mullapudi_automatically_2016}.
Automatic scheduling promises a realistic path towards integration into high-level
systems like SciPy \cite{virtanen_scipy_2020} or TensorFlow \cite{abadi_tensorflow_2016}.
Whereas sparse tensor compilers have made performance engineers more productive, our automatic scheduler will make
end users more productive.

We present, to the best of our knowledge, the first automatic scheduler for
asymptotic decision making in sparse tensor programs. At its core is an
asymptotic cost model that can automatically analyze and rank the complexity of
sparse tensor schedules over all possible inputs to the program. We introduce a
general language for describing schedules, the Protocolized Concrete Index
Notation (abbreviated CIN-P).  Autoschedulers often make high-level decisions
before considering fine-grained implementation
details\cite{anderson_efficient_2021}.  Our autoscheduler ignores
constant-factor optimizations such as sparse format implementation details,
cache blocking, or parallelization. We focus only on novel asymptotic concerns,
such as loop fusion, loop reordering, and format selection.  Our method is able
to detect cases where one program performs strictly less work than another
across all input patterns, up to constant factors.  The result is a frontier of
asymptotically undominated programs, one of which being the best choice for
any given input, up to constant factors. Programs in the (manageably small)
frontier may be later embellished with constant-factor optimizations.

Tensor kernels are usually static from run to run. We therefore design for an offline use case,
where the autoscheduler runs only once per kernel and is given no
information about the sparsity patterns of the inputs. In contrast, online
autoschedulers execute at runtime, running once per input pattern and making use
of the pattern to select an appropriate specialized implementation just-in-time.
Both regimes are important. While the runtime of online autoschedulers
competes with the optimizations they deliver, the runtime of offline
autoschedulers only competes with the equivalent developer effort and salary required to
write a schedule.  Offline autoschedulers might run on a
dedicated server as new schedules for kernels are requested by users, and new
schedules could ship with each update to the tensor compiler. The TACO web
scheduling tool has recorded only 2758 distinct tensor programs since 2017\footnote{According to private correspondence with the author of the tool.}.
Because sparse tensor programs are often small (only a few tensors and indices) and the offline scheduling use case affords an
extensive amount of time to produce schedules, exponential-time solutions to
difficult offline scheduling problems are within reach.

Our contributions are as follows:
\begin{itemize}
  \item We define a language (CIN-P) for the implementation of sparse tensor programs at
  a high level, specifying the loop structure, temporary tensors, and tensor
  formats. Our language separates the storage formats from how they should be
  accessed (the protocol).
  \item We model asymptotic complexity using abstract set expressions. We
  describe algorithms to derive the modeled complexity cost of CIN-P schedules and
  determine when one complexity dominates another.
  \item We use our cost model to write an asymptotic autoscheduler for CIN-P
  sparse programs. We enumerate equivalent programs of minimal loop nesting
  depth, filter these programs to the asymptotic frontier, and use a novel
  algorithm to automatically insert workspaces for transpositions and
  reformatting. We demonstrate that our asymptotic frontier is often several
  orders of magnitude smaller than the minimum depth frontier.
  \item We evaluate our approach on the subset of CIN-P programs supported by
  the TACO tensor compiler, demonstrating performance improvements of several
  orders of magnitude over the default schedules. 
\end{itemize}

\section{Background}
Tensor compilers provide a mechanism to automatically generate efficient code
for simple loop programs that operate on tensors, or multidimensional arrays
\cite{kotlyar_compiling_1997, kotlyar_relational_1997-1, kotlyar_relational_1999, bik_automatic_1994,bik_compilation_1993, pugh_sipr_1999, strout_approach_2016, arnold_data-parallel_2011,venkat_loop_2015, kjolstad_tensor_2017, kjolstad_tensor_2019, chou_automatic_2020, chou_format_2018, henry_compilation_2021, tian_high-performance_2021}.
Sparse tensor compilers are
specialized for the case where tensors are mostly zero and only nonzero elements
are stored, making the problem especially complex.  In addition to loop ordering
problems of the dense case, the sparse code must also iterate over compressed
representations of the input. Sparse matrix representations have a long history
of study, and are typically specialized to the kernel to be executed. As an
example, consider TACO, a recently popular sparse tensor compiler that has
inspired several lines of inquiry due to it's simplified abstractions for sparse
compilation \cite{kjolstad_tensor_2017,
kjolstad_tensor_2019}. Such investigations include modifications to the compiler itself
\cite{chou_automatic_2020, chou_format_2018, henry_compilation_2021} or separate
implementations such as COMET \cite{tian_high-performance_2021} or the MLIR
sparse tensor dialect.  TACO simplifies compilation of sparse tensor programs by
considering each dimension separately. A type system is used to specify whether each dimension
is to be compressed (and therefore iterated over), or dense (and accessed with
direct memory references). Workspaces, or temporary tensors, may be introduced
to hold intermediate results. The loop ordering, workspaces, and sparse formats
of the inputs together form the high-level description, a \textbf{schedule},
from which TACO generates code. 

Different schedules may have different asymptotic effects that depend on the
sparsity patterns of the input. TACO generates code that takes advantage of the
properties that $a \cdot 0 = 0$ and $a + 0 = a$, so only nonzero values need to
be processed, and when tensors are multiplied, only their shared nonzero values 
need to be recorded. The former rule is more important, since we can avoid
computing $a$ in the first place. Workspaces may be inserted to cache
intermediate results to avoid redundant computation, provide an effective buffer
to avoid asymptotically expensive operations on sparse tensor formats, or
perform filtering steps by exposing intermediate zero values early in a
computation. The order in which sparse loops are evaluated can have asymptotic
effects as well, since sparse outer loops may act as filters over their
corresponding inner loops.

\section{Motivating Examples}\label{sec:motivatingexample}

As an example, notice that the three main algorithms for sparse matrix-matrix
multiply (SpGEMM) can be thought of as differing only in their loop ordering. We
write SpGEMM as
\[
  A_{ij} = \sum_k B_{ik} \cdot C_{kj}.
\]
The ``Inner Products'' approach processes this expression as a set of loops
nested in $i, j, k$ order, with the inner loop performing sparse inner products, merging nonzeros in rows of $B$ and columns of $C$ \cite{mcnamee_algorithm_1971}. Even though we only need to multiply the shared nonzeros, merging the lists has a runtime proportional to their length, regardless of how many nonzeros they share. The improved ``Outer Products'' algorithm loops in $k, i, j$ order, scattering
into $A$ and only iterating through shared nonzeros\cite{buluc_representation_2008}. Setting $k$ to be the outer
loop ensures that all nonzero $i$ and $j$ encountered in inner loops share a value of $k$ and correspond to
necessary work. Gustavson's algorithm iterates in $i, k, j$ order, representing
a compromise between the two approaches that avoids the need to scatter into a
two-dimensional output \cite{gustavson_two_1978}. It should be noted that scattering operations incur
asymptotic costs as well depending on the format used to store the output.

Loop fusion can be critical for sparse programs like sampled dense-dense matrix
multiplication, written as
\[
  A_{ij} = \sum_k B_{ik} \cdot C_{kj} \cdot D_{ij}
\]
where $B$ and $C$ are now dense, but $D$ is a sparse matrix. If we process this
kernel as a dense matrix multiply and then a sparse mask operation, like $A_{ij}
= w_{ij} \cdot D_{ij} \where w_{ij} = \sum_k B_{ik} \cdot C_{kj}$, then the
runtime is $O(I \cdot J \cdot K)$, whereas if we instead process all three operands at
once, the runtime is $O(nnz(D) \cdot K)$.

On the other hand, inserting temporaries and avoiding loop fusion can be
critical for kernels like the three-way pointwise sparse matrix product,
\[
  A_{ij} = B_{ij} \cdot C_{ij} \cdot D_{ij}.
\]
Inserting a temporary like 
\[A_{ij} = w_{ij} \cdot D_{ij} \where w_{ij} = B_{ij} \cdot C_{ij}
\]
will avoid reading rows of $D$ when the product $B_{ij} \cdot C_{ij}$ produces
empty rows.

\section{Protocolized Concrete Index Notation}

In order to make our descriptions of sparse tensor algebra implementations more
precise, we introduce the protocolized concrete index notation, an extension to
the concrete index notation of \cite{kjolstad_tensor_2019}. Our notation starts
with the tensors themselves. A more detailed description of TACO-style formats
is given in \cite{chou_format_2018}. A \textbf{rank}-$r$ tensor of
\textbf{dimension} $I_1, ..., I_r$ maps $r$-tuples of integers $i \in 1:I_1
\times ... \times 1:I_r$ to values $v$. Each position in the tuple is referred
to as a \textbf{mode}. We represent sparse tensors using trees where each node
$i_1, ..., i_l$ at \textbf{level} $l$ in the tree represents a slice $T_{i_1,
..., i_l, :, ..., :}$ of the tensor that contains at least one nonzero. Each
level is stored in a particular format. If the format is \textbf{uncompressed}
($\farray$), then for each $i_1, ..., i_{l - 1}$ represented by the previous
level, we store an array of all possible children $1:I_l$. If the
format is \textbf{compressed} ($\flist$), then we only store a list of nonzero
children (and their locations). If the format is a \textbf{hash table}
($\fhash$), we store the same information as the list format, but we use a hash
table, enabling random access and insertion. We abbreviate our formats with
their first letter and specify them as superscripts, read from left to right
corresponding to the top down to the bottom of the compressed tensor tree. A
matrix stored in the popular \textbf{CSR} format (rows are stored at the top
level in an array, and columns as a list in the bottom level), would be written
as $A^{\farray\flist}$.  If the modes are to be stored in a different order, we
write the permutation next to the format, so \textbf{CSC} format (where columns
are stored in the top) would be written as $A^{\farray(2)\flist(1)}$.

The following paragraph summarizes concrete index notation, described more
formally in \cite{kjolstad_tensor_2019}.
We use \textbf{index variables} to specify a particular element of a tensor.
Tensors may be \textbf{accessed} by index variables $i$ as $A_i$. We can
combine accesses into \textbf{index expressions} with function calls, such as the calls to $+$ and $\cdot$ in
$B_{ij} + 2 \cdot C_{k}$. An \textbf{assignment} statement writes to an element of a
tensor, and takes the form $A_{i...} = expr$, where $expr$ is an index
expression. An \textbf{increment} statement updates an element of a tensor,
taking the form $A_{i...} \mathrel{f}= expr$ where $f$ is a binary operator such as $+$ or $\cdot$ and roughly meaning $A_{i...} = f(A_{i...},
expr)$, although we disallow the left hand side tensor from appearing on the
right hand side of assignment or increment statements. The assignment statement
``returns'' the tensor on it's left hand side, which can be used by a
\textbf{where} statement. The statement $consumer \where producer$ evaluates the
statement $producer$, and makes the tensor it returns available for use in the
scope of $consumer$, returning the tensor returned by $consumer$. The
\textbf{forall} statement $\forall_{i...} body$ evaluates $body$ over all
assignments to the indices $i...$ and returns the tensor returned by $body$.
However, when operands are sparse, we can skip some of these evaluations.

Adding to the existing concrete index notation, we introduce the notion of a
\textbf{protocol}, used to describe how an index variable should interact with
an access when that variable is quantified.  The \textbf{step} ($\pstep$)
protocol indicates that the forall should coiterate over a list of nonzeros of
the corresponding tensor, and substitute the default tensor value $0$ into the
body for the other values. The \textbf{locate} ($\plocate$) protocol indicates
that the forall should ignore this access for the purposes of determining which
values to coiterate over.  The list format only supports the step protocol, and
the array format only supports the locate protocol, but the hash format supports
both. We use separate protocols for writes. We say that a write is an
\textbf{append} ($\pappend$) protocol when we can guarantee that writes to that
mode and all modes above it in the tensor tree will occur in lexicographic
order. We say that a write is an \textbf{insert} ($\pinsert$) protocol
otherwise. In effect, append and insert mirror the flavor of step and locate. We
abbreviate our protocols with their first letter and specify them like functions
surrounding indices in our access expressions, meant to specify how each mode of
the tensor should be accessed. If we mean to read the mode indexed by $i$ using
a step protocol and the mode indexed by $j$ using a locate protocol, we would
write this as $A_{\pstep(i)\plocate(j)}$.

It helps to have some examples of protocolized concrete index notation and the kind
of code it might generate. We start with our matrix multiplication kernel
$A_{ij} = \sum_k B_{ik} \cdot C_{kj}$.  Let's assume we have sparse matrices
$B^{\flist\flist}$ and $C^{\flist\flist}$, and wish to produce
$A^{\flist\flist}$ (i.e. all matrices are stored in DCSR format, both dimensions
are stored compressed, similar to a list of lists).  The inner products matrix
multiply could be written as
\begin{multline*}
  \big(\forall_{ijk} A^{\flist\flist}_{\pappend(i)\pappend(j)} \pluseq B^{\flist\flist}_{\pstep(i)\pstep(k)} \cdot {C'}^{\flist\flist}_{\pstep(j)\pstep(k)}\big) \where \\
  \big(\forall_{jk} {C'}^{\flist\flist}_{jk} = C^{\flist\flist}_{kj}\big)
\end{multline*}
where the producer side of the where statement transposes $C$ into a workspace
$C'$ so that the tree order in the consumer side agrees with the quantification
order. Ignoring the transpose, the corresponding pseudocode would look like:
\begin{algorithmic}
  \For{$i \in B$}
    \For{$j \in C'$}
      \For{$k \in B_i \cup {C'}_j$}
        \If{$B_{ik} \neq 0 \wedge {C'}_{jk} \neq 0$}
          \State{$A_{ij} \gets A_{ij} + B_{ik} \cdot {C'}_{jk}$}
        \EndIf
      \EndFor
    \EndFor
  \EndFor
\end{algorithmic}
This code will iterate over $B_i \cup {C'}_j$, even though it only needs
to iterate over $B_i \cap {C'}_j$.

To improve the situation, we might choose to use the outer products algorithm, 

\begin{multline*}
  \big(\forall_{ij} A^{\flist\flist}_{ij} = {A'}^{\fhash\fhash}_{ij}\big) \where \\ 
    \Big(\big(\forall_{kij} A^{\fhash\fhash}_{\pinsert(i)\pinsert(j)} \pluseq {B'}^{\flist\flist}_{\pstep(k)\pstep(i)} \cdot C^{\flist\flist}_{\pstep(k)\pstep(j)}\big) \where \\
    \big(\forall_{ki} {B'}^{\flist\flist}_{ki} = B^{\flist\flist}_{ik}\big)\Big)
\end{multline*}

where we have used a temporary hash format to handle the random accesses to $A$
and a transposition of $B$ to access $k$ first. The resulting code would look like
\begin{algorithmic}
  \For{$k \in B' \cup C$}
    \If{${B'}_{k} \neq 0 \wedge C_k \neq 0$}
      \For{$i \in B'_k$}
        \For{$j \in C_k$}
          \State{$A_{ij} \gets A_{ij} + B'_{ki} \cdot C_{kj}$}
        \EndFor
      \EndFor
    \EndIf
  \EndFor
\end{algorithmic}

In this version, we have avoided repeating the filtering step for every $i$ and
$j$. Unfortunately, this version introduces a two-dimensional scatter, which can
be expensive. Instead, we may choose to use Gustavson's algorithm, written as:

\begin{multline*}
    \forall_i \Big(
    \big(\forall_{j} A^{\flist\flist}_{\pappend(i)\pappend(j)} = w^{\fhash}_{\pstep(j)}\big) \where \\
    \big(\forall_{kj} w^{\fhash}_{\pinsert(j)} \pluseq B^{\flist\flist}_{\pstep(i)\pstep(k)} \cdot C^{\flist\flist}_{\pstep(k)\pstep(j)}\big)\Big)
\end{multline*}

This is our first example with a quantified where statement. It expands to:

\begin{algorithmic}
  \For{$i \in B$}
    \State{$w \gets 0$} \Comment{Initialize $w$}
    \For{$k \in B_i \cup C$}
      \If{$B_{ik} \neq 0 \wedge C_k \neq 0$}
        \For{$j \in C_k$}
          \State{$w_{j} \gets w_{j} + B_{ik} \cdot C_{kj}$}
        \EndFor
      \EndIf
    \EndFor
    \For{$j \in w$}
      \State{$A_{ij} \gets w_{j}$}
    \EndFor
  \EndFor
\end{algorithmic}

This form of the algorithm is a practical improvement over the outer products
formulation because the workspace is one dimensional, meaning that it can be
implemented with a dense vector rather than a hash table. Note that workspaces
are initialized just before executing the where statement that returns them.

\section{Cost Modeling}

In this section, we formalize prior intuitions by describing a language for
sparse asymptotic complexity. We break up the runtime of sparse workloads into
individual constant-time tasks, represented by the index variable
values in the scope that the task executes. For example, when $i = 3$, $j = 7$,
and $k = 2$, and we execute the body of the inner loop of $\forall_{ijk} A_{ij} \pluseq
B_{ik} \cdot C_{kj}$, the multiplication, addition, and access expressions incur a
cost represented by the task $(3, 7, 2)$. There are two kinds of tasks in a
sparse program, computation tasks that cover the numerical body of the loop, and
coiteration tasks that cover the cost of iterating over multiple sparse input
tensor levels (not all iterations lead to compute). We can then use a modified
set-building notation to describe the set of tasks that each computation incurs.
For example, computing the expression $A_{ij} = B_{ij} \cdot C_{ij}$
would incur a coiteration cost proportional to the size of the set \[
  \{(i, j) \in I \times J \mid B_{ij} \vee C_{ij}\},
\] and a computation cost proportional to the size of the set \[
  \{(i, j) \in I \times J \mid B_{ij} \wedge C_{ij}\}.
\]
where we overload our notation for sparse tensors as Boolean predicates
for whether the corresponding entry of the tensor is nonzero, and
we overload our notation for dimensions as the set of all valid indices into the
corresponding modes. We sometimes omit these dimensions for brevity.

As another quick example, the code to compute the outer product $A_{ij} = b_i * c_j$ of a sparse
vector $b$ and a dense vector $c$ would have a complexity
proportional to ($|\{(i, j) \in I \times J \mid b_i\}| = J * \nnz(b)$). Notice
that $j$ is unconstrained here since $c$ is dense.

Having defined our new notation, we precisely characterize the differences between
our example approaches to sparse matrix multiplication in Figure
\ref{fig:thegemmfigure}.  Notice that the $k$ loop of the inner products algorithm
will incur the largest cardinality task set among any of the sets listed, regardless of
the particular patterns of $A$ or $B$. This is because it iterates over the
nonzeros in both $A$ and $B$ for each $i$, $j$ pair, regardless of how many $k$
values are shared, essentially repeating a filtering step. Gustavson's algorithm
and the Outer Products algorithm perform this filtering higher in the loop nest.

\begin{figure}

  {\large Inner Products SpGEMM}
  \begin{align*}
    &\forall i &\{[i] \mid \exists_{k_1} B_{ik_1}\} \cup\\
    &\quad \forall j &\{[i, j] \mid \exists_{k_1,k_2} B_{ik_1} \wedge C_{k_2j}\}\cup\\
    &\quad\quad  \forall k &\{[i, j, k] \mid B_{ik} \vee C_{kj}\}\cup\\
    &\quad\quad\quad   A_{ij} \pluseq B_{ik} \cdot C_{kj} &\{[i, j, k] \mid B_{ik} \wedge C_{kj}\}
  \end{align*}
  {\large Gustavson's SpGEMM}
  \begin{align*}
    &\forall i &\{[i] \mid \exists_{k_1} B_{ik_1}\}\cup\\
    &\quad \forall k &\{[i, k] \mid \exists_{j_1} B_{ik} \vee C_{kj_1}\}\cup\\
    &\quad\quad  \forall j &\{[i, j, k] \mid B_{ik} \wedge C_{kj}\}\cup\\\
    &\quad\quad\quad   A_{ij} \pluseq B_{ik} \cdot C_{kj} &\{[i, j, k] \mid B_{ik} \wedge C_{kj}\}
  \end{align*}
  {\large Outer Products SpGEMM}
  \begin{align*}
    &\forall k &\{[k] \mid \exists_{i_1, j_1} B_{i_1k} \vee C_{kj_1}\}\cup\\
    &\quad \forall i &\{[i, k] \mid \exists_{j_1} B_{ik} \wedge C_{kj_1}\}\cup\\
    &\quad\quad \forall j &\{[i, j, k] \mid B_{ik} \wedge C_{kj}\}\cup\\
    &\quad\quad\quad A_{ij} \pluseq B_{ik} \cdot C_{kj} &\{[i, j, k] \mid B_{ik} \wedge C_{kj}\}
  \end{align*}
  \caption{Some example implementations of SpGEMM and the corresponding
  asymptotic costs.}\label{fig:thegemmfigure}
\end{figure}

\begin{figure}
  {\large Fused SDDMM}
  \begin{align*}
    &\forall i &\{[i] \mid \exists_{k_1} D_{ik_1}\}\cup\\
    &\quad \forall j &\{[i, j] \mid D_{ij}\}\cup\\
    &\quad\quad  \forall k &\{[i, j, k] \mid D_{ij}\}\cup\\
    &\quad\quad\quad   A_{ij} \pluseq B_{ik} \cdot C_{kj} \cdot D_{ij} &\{[i, j, k] \mid D_{ij}\}
  \end{align*}
  {\large Non-Fused SDDMM}
  \begin{align*}
    &\quad\forall i &\{[i] \mid \exists_{k_1} D_{ik_1}\}\cup\\
    &\quad\quad \forall j &\{[i, j] \mid D_{ij}\}\cup\\
    &\quad\quad\quad   A_{ij} \pluseq w_{ij} \cdot D_{ij} &\{[i, j] \mid D_{ij}\}\\
    &\where\\
    &\quad\forall i &\{[i]\}\cup\\
    &\quad\quad \forall j &\{[i, j]\}\cup\\
    &\quad\quad\quad  \forall k &\{[i, j, k]\}\cup\\
    &\quad\quad\quad\quad   w_{ij} \pluseq B_{ik} \cdot C_{kj} &\{[i, j, k]\}
  \end{align*}
  \caption{Some example implementations of SDDMM and the corresponding
  asymptotic costs.}\label{fig:thesddmfigure}
\end{figure}

\subsection{Asymptotic Domination}
Implicit in our discussion until now has been the notion of asymptotic
domination among task sets. Usually, we say that a runtime $f$ asymptotically
dominates a runtime $g$ with respect to a sequence of inputs $x_n$ if for every
$c > 0$, there exists $n_0$ such that $f(x_n) > c \cdot g(x_n)$ for all $n > n_0$.
However, this definition relies on the choice of inputs for the two algorithms.
For instance, when computing the expression $a_{ij} = b_{ij} \cdot c_{ij} \cdot d_{ij}$,
one might compute $b_{ij} \cdot c_{ij}$ first, avoiding the need to traverse $d$
when $b$ and $c$ are disjoint. However, a similar argument could be made for
grouping $c$ and $d$ first. 

We say that a task set $f$ is asymptotically dominated by another task set $g$
when $f$ is contained in $g$ but $g$ is not contained in $f$. 

While tasks may be defined by different numbers of index variables in different
orders, we would like the task set $\{[i, j, k] \mid A_{ijk}\}$ to contain the task
set $\{[j, i] \mid \exists_k A_{ijk}\}$, for example. We therefore interpret each
task as a tuple of its indices, and as a shorthand for all tuples of subsets of
its indices and permutations thereof. Thus, $\{[i, j] \mid \exists_k A_{ijk}\}$ is
a shorthand for
\begin{align*}
  &\{(i, j) \mid \exists_k A_{ijk}\} \cup \\
  &\{(j, i) \mid \exists_k A_{ijk}\} \cup \\
  &\{(i) \mid \exists_{jk}  A_{ijk}\} \cup \\
  &\{(j) \mid \exists_{ik} A_{ijk}\} \cup \\
  &\{() \mid \exists_{ijk} A_{ijk}\}.
\end{align*}

As more information about the inputs is made known to the autoscheduler, more
detailed analyses can be performed to construct proofs that one kernel may be
better than another on a class of inputs.  Our method will prove task set
containment whenever it is possible to do so without any knowledge of the data
pattern. While our set containment metric is not necessarily equivalent to
asymptotic domination, we show that our analysis explains the asymptotic
behaviors of kernels previously studied by numerical analysts, and that our
analysis can effectively prune the kernel search space down to a manageably
sized frontier.

At this point, our task sets have been defined as sets of tuples with Boolean
predicates that contain only existential quantifiers, conjunctions, and
disjunctions. We can normalize these sets to a standard ``union of conjunctive
queries'' form \cite{chandra_optimal_1977, kolaitis_conjunctive-query_2000, konstantinidis_scalable_2013}. A conjunctive query is defined as the set of satisfying variable
assignments (expressed as a tuple of values) to an existentially quantified
conjunction of Boolean functions on the variables. We write conjunctive queries as
\[
\{(i_1, i_2...) \mid \exists_{j_1, j_2...} A_{k_1, k_2, ...} \wedge B_{l_1, l_2, ...} ...\}
\]
The tuple $(i_1, i_2, ...)$ is referred to as the \textbf{head}, $A_{k_1, k_2,
...}$, and $B_{l_1, l_2, ...}$ are \textbf{clauses}, and our tensor variables
$A, B, ...$ themselves are \textbf{predicates}. In our clauses, the predicates
may be indexed by any combination of quantified or head variables.

It has been shown that a conjunctive query $P$ is contained in another query $Q$
if and only if we there exists a homomorphism $h$ from the variables of $Q$ to
those of $P$. A variable mapping $h : Q \to P$ is a homomorphism if applying $h$
to the head of $Q$ gives the head of $P$ and if every clause $A_{k_1, k_2, ...}$
in $Q$ has a corresponding clause $A_{h(k_1), h(k_2), ...}$ in $P$.

Furthermore, a union of conjunctive queries $P_1, P_2, ...$ is contained in a
union of conjunctive queries $Q_1, Q_2, ...$ if each $P_i$ is contained in at
least one $Q_j$. This implies a fairly straightforward algorithm for checking
containment.  For each conjunctive query in $P$, we attempt to find a
conjunctive query in $Q$ that contains it.  We determine this containment by
performing a backtracking search for homomorphisms, maintaining the partial
homomorphism as we process each clause in turn. At each clause $P_i$, we make a 
choice of which clause $Q_j$ will cover it. If we run into a variable conflict
in the homomorphism, we backtrack and try a different $Q_j$. This algorithm may
take exponential time, but recall that our tensor programs are of small,
constant size.

Using De Morgan's law, we can convert conjunctions of disjunctions into
disjunctions of conjunctions, and we can convert set expressions over
disjunctions into unions of set expressions. We can move conjunctions and
disjunctions inside of existential quantifiers by renaming the quantified
variables. Thus, we can normalize all of our task sets into unions of
conjunctive queries and determine containment.

\subsection{Dimensions and Lazy Tasks}
The head variables in our task sets need a notion of dimension. Without
including dimension into our predicates, a program that iterates over the
entirety of some index variable $i$ would have a task set representation $\{[i] \mid true\}$,
an infinitely large set! This is not a problem for existentially
quantified variables since they do not contribute to the size of our task sets.
Formally, we constrain the dimensions of head variables by interpreting each
dimension itself as a predicate. Our task set would then be $\{[i] \mid I_i\}$.
Because tensor predicates also imply that their index variables are in bounds,
we must represent that implication explicitly as well. For example, iterating
over an $I \times J$ matrix $A$ incurs a task set $\{[i, j] \mid A_{ij} \wedge I_i \wedge J_j\}$.
To avoid expanding our predicates with these extra clauses, we
instead move the representation of dimension to the head of the query itself.
Thus, we write our previous expression as $\{[i \in I, j \in J] \mid A_{ij}\}$.
We then require that any homomorphisms respect the dimensionality of the head
variables.

Recall that the each task is semantically expanded into tuples of all subsets of
it's indices, and all permutations thereof. In order to avoid exponential
increases in the input size of what is already an exponential time containment
algorithm, we represent this expansion lazily. Given two task sets $P = \{[i_1,
i_2, ...] \mid A\}$ and $Q = \{[j_1, j_2, ...] \mid B\}$, we can say that $P
\subseteq Q$ if our algorithm can find a homomorphism $h$ from $Q$ to $P$ where
$\{i_1, i_2, ...\} \subseteq \{h(j_1), h(j_2), ...\}$. This slight modification
to our definition of homomorphisms allows us to avoid representing all subsets
and permutations of each task explicitly. Normally, our search for homomorphisms
would take the mapping between head variables to be a given. Now, we must
perform such a search for each mapping of the head of $P$ to a subset of the
head of $Q$. Recalling that our homomorphisms must respect the dimension of head
variables, we can thankfully constrain the number of head variable mappings that
we search through. The combination of representing dimensionality in the head of
the clause and representing powersets of tasks lazily allows us to reduce the
runtime of our algorithm significantly.

\subsection{Sunk Costs and Assumptions}
Set containment analysis can be quite strict as written. For example, the set
containment metric prefers the outer products algorithm over Gustavson's
algorithm when all matrices are DCSR, even though both algorithms are quite
practical. The reason for this is that when $C$ is all zero, Gustavson's
algorithm still iterates over the entirety of $B$ whereas the outer products
algorithm doesn't. Practitioners would say that this case is unrealistic, since
we shouldn't expect operands to be all-zero, and even if this case did occur,
simply iterating over the nonzeros of $B$ isn't a big deal. We agree, and extend
our analysis to reflect common sunk costs and assumptions. Assume we are to
compare two costs $P$ and $Q$. If there are any sunk costs $S$ (such as the
linear-time costs of reading all the inputs), we add those costs to our queries
and instead compare $P \cup S$ to $Q \cup S$. If there are any assumptions $A$,
we add those assumptions to the predicates of our queries and compare $\{P \mid A\}$
to $\{Q \mid A\}$. Throughout the rest of the paper, we will take the time to read
sparse (not dense) inputs and the time to iterate over any single dimension as
sunk costs. We will also assume that all sparse inputs are nonempty.

\subsection{Building a Frontier}
Recall that different implementations may perform differently depending on the
input patterns, so our cost model is unable to identify a single best
implementation. Instead, we can use our cost model to produce a frontier of
kernels whose runtimes do not strictly dominate those of any other kernels. Our
algorithm starts with an empty frontier and processes each kernel in turn. If
the current kernel dominates a kernel in the frontier, we discard it. Otherwise,
we add the current kernel to the frontier and remove any other kernels that
dominate it. This algorithm avoids a strictly quadratic number of containment
checks by only comparing programs to the current frontier instead of the
universe, but we cannot guarantee any bounds on the intermediate size of the
frontier so this improvement is only heuristic.

\begin{figure}
  \caption{Our algorithm for producing an asymptotic frontier. In practice, the
  size of the frontier does not grow too large, and the algorithm performs much
  less than the worst case quadratic number of asymptotic comparisons.}
  \begin{algorithmic}
    \State{$frontier \gets []$}
    \For{$(S, P) \in universe$}
      \For{$(T, Q) \in frontier$}
        \If{$P \subseteq Q$ and not $Q \subseteq P$}
          \State{Add $(S, P)$ to frontier if not added already}
          \State{Remove $(T, Q)$ from frontier}
        \EndIf
      \EndFor
    \EndFor
  \end{algorithmic}
\end{figure}

Since we perform asymptotically more containment checks than we have possible
programs, it can be helpful to preprocess the runtimes to reduce the complexity
of the task sets and allow the containment checks to run faster. We make three
simplifications.

First, we add our sunk costs and assumptions to each runtime and normalize the
resulting terms before running our frontier algorithm.

Second, recall that we represent our task sets using a union of conjunctive
queries $Q_1 \cup Q_2 \cup ...$. If it happens that for some $i$ and $j$, $Q_i
\subseteq Q_j$, then we can leave $Q_i$ out.
\[
  Q_1 \cup ... \cup Q_i \cup ... \cup Q_j \cup ... = Q_1 \cup ... \cup Q_{i - 1} \cup Q_{i + 1} \cup ... \cup Q_j \cup ...
\]
By checking for containment among all pairs of conjunctive terms in the union,
we can simplify the description of the task set to a unique union of undominated
conjunctive queries.

Third, we iteratively check for simplifications in each conjunctive query by
leaving each term out and checking for containment. For example, if we
have a conjunctive query $\{(i...) \mid C_1 \wedge C_2 \wedge ... \}$, we check
\[
\{(i...) \mid C_1 \wedge C_2 \wedge ... \} = \{(i...) \mid C_1 \wedge
... \wedge C_{j - 1} \wedge C_{j + 1} \wedge ... \}
\]
for each $j$ in turn. If we do manage to simplify our query, we repeat the
process until we cannot make the query any smaller. Unlike our previous
optimization, we cannot guarantee whether the result of this optimization is
unique.  However, we find that it heuristically simplifies our terms and reduces
our runtime considerably.

\subsection{Automatic Asymptotic Analysis}

With the goal of using our asymptotic analysis as part of an automated
scheduler, we describe an algorithm to return the asymptotic complexity of an
input tensor program in protocolized sparse concrete index notation.

Our algorithm performs an abstract interpretation over each node of the program,
using a Boolean predicate to describe the set of iterations currently being
executed (referred to in our pseudocode as the ``guard''). We also construct
Boolean predicates that represent the nonzero locations written to during the
course of executing the program (referred to as the ``state''). Because tensors
may be written and read by index variables of different names, we rename the
variables in each predicate after the mode of the tensor they represent. Thus,
when we write $state(A_i) \gets state(A_{i...}) \vee \exists_j B_{ij}$, we mean to
update our state's predicate for $A$ to add any nonzero locations in the pattern
$\exists_j B_{ij}$.

We start our traversal at the topmost node with no bound variables, a guard set
to true, and a state filled default inputs ($state(A_{i...}) = A_{i...}$ for all
inputs $A$).

When we encounter a forall node over an index $i$, we collect all of the
accesses in the body that access $i$ with a step protocol. Since our sparse
program will need to coiterate through these tensors, we output the task set
$\{[j...] \mid \exists_{k...} A_{l...}\}$ for each access $A_{l...}$, where $j$
is the set of bound variables and $k \gets l \setminus j$. Each access might be
zero or nonzero. As it coiterates over all nonzeros in each of the tensors, our
sparse code will only execute the body in cases where there are nonzero operands
that need processing. Thus, our abstract interpretation iterates over every
combination of zero or nonzero for each tensor, substituting the zeros into the
body and simplifying before recursing. When we recurse, we use the zeroness or
nonzeroness of each access as a guard on the set of iterations that the
recursive call corresponds to. If no tensors access $i$ with step protocol, then
we can simply recurse on the body after adding $i$ to the set of bound
variables.

When we encounter a where node, we add a zero-initialized workspace to the state
of the producer before processing the producer side, and we make that new tensor
state available when we process the consumer side.

When we encounter an assignment statement $A_{i...} \pluseq ...$, we output
$\{[bound...] \mid guard\}$ to reflect the work performed by this statement. We
also update the state of $A{i...}$ to add in the writes represented by $guard$.

\begin{figure}
  \caption{Our algorithm for analyzing the complexity of programs in
  protocolized concrete index notation. The $tasks$ variable is a global that
  records the total complexity, and the $state$ variable is a global dictionary
  that we use to hold the locations of nonzeros written to each tensor over the
  course of the program.}\label{alg:complexity}
\begin{algorithmic}
\State{$tasks \gets \emptyset$}
\State{$state \gets \text{predicates initialized with input tensors}$}
\Procedure{Complexity}{$node$, $bound$, $guard$}
  \If{$node$ matches $\forall_i body$}
    \State{$steppers \gets$ all step protocol accesses of $i$ in $body$}
    \State{$bound' \gets (bound..., i)$}
    \State{$cases \gets \{body\}$}
    \For{$A_{j...} \in steppers$}
      \State{expand $cases$ by substituting $A_{j...} \to 0$}
      \State{$k \gets (j \setminus bound')$}
      \State{$iters \gets \{[bound'...] | guard \wedge \exists_{k...} state(A_{j...})\}$}
      \State{$tasks \gets tasks \cup iters$}
    \EndFor
    \For{$body' \in cases$}
      \State{simplify $body'$ using zero-annihilation}
      \State{$guard' \gets guard$}
      \For{$A_{j...} \in steppers$}
        \If{$A_{j...}$ is in $body'$}
          \State{$k \gets (j \setminus bound')$}
          \State{$guard' \gets guard' \wedge \exists_{k...} state(A_{j...})$}
        \EndIf
      \EndFor
      \State{\Call{Complexity}{$body'$, $bound'$, $guard'$}}
    \EndFor
  \ElsIf{$node$ matches $cons \where prod$}
    \State{$state[result(prod)] \gets 0$}
    \State{\Call{Complexity}{$prod$, $bound$, $guard$}}
    \State{\Call{Complexity}{$cons$, $bound$, $guard$}}
  \ElsIf{$node$ matches $A_{i...} \pluseq ...$}
    \State{$computes \gets \{[bound...] \mid guard\}$}
    \State{$tasks \gets tasks \cup computes$}
    \State{$state(A_{i...}) \gets state(A_{i...}) \vee guard$}
  \EndIf
\EndProcedure
\end{algorithmic}
\end{figure}

Our algorithm is summarized in Figure \ref{alg:complexity}.
A benefit of our algorithm to analyze complexity is that it will extend to
alternate fill values and operators which are not $+$ or $\cdot$, extensions to the
TACO compiler explored in \cite{henry_compilation_2021}. Nothing in our
algorithm is specific to the choice of $0$ or the operators we have chosen in
our examples.

\section{Autoscheduling}\label{sec:schedule}

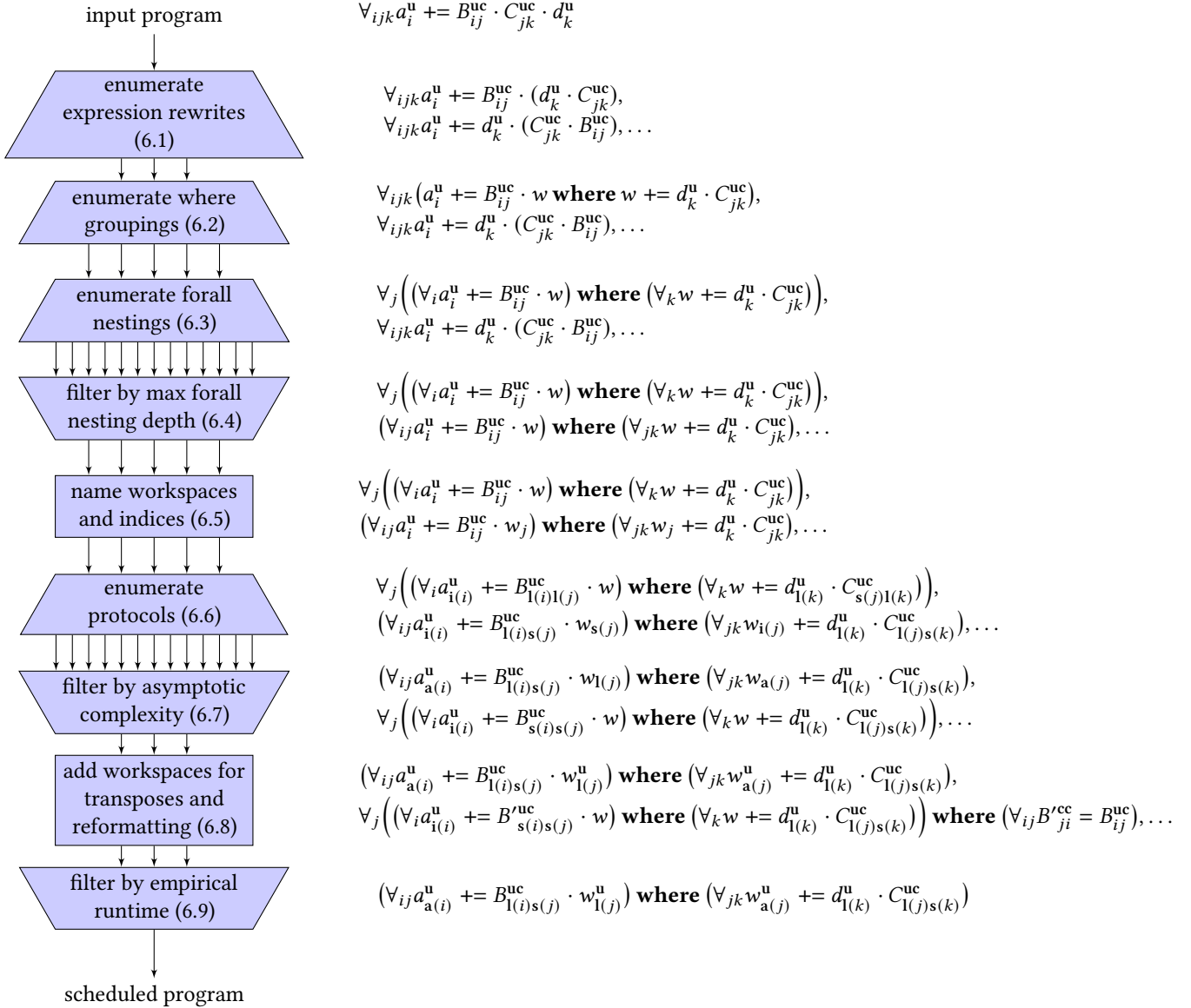
\begin{figure*}
\tikzstyle{choice} = [trapezium, draw, trapezium angle=60, fill=blue!20, 
    text width=8em, text badly centered, inner sep=3pt]
\tikzstyle{filter} = [trapezium, draw, trapezium angle=-60, fill=blue!20, 
    text width=8em, text badly centered, inner sep=3pt]
\tikzstyle{pass} = [rectangle, draw, fill=blue!20, 
    text width=8em, text badly centered, inner sep=3pt]
\tikzstyle{program} = [rectangle,
    text width=8em, text badly centered, inner sep=3pt]
\tikzstyle{example} = [rectangle,
    text width=40em, inner sep=3pt]
\tikzstyle{line} = [draw, -latex']
    
\begin{tikzpicture}[node distance = 1.5cm, auto]
    \node [program] (init) {input program};
    \node [example, right=of init.east] {
      $\forall_{ijk} a^{\farray}_{i} \pluseq B^{\farray\flist}_{ij} \cdot C^{\farray\flist}_{jk} \cdot d^{\farray}_{k}$
    };
    \node [choice, below of=init] (churn) {enumerate expression rewrites (\ref{sec:schedule:churn})};
    \node [example, right=of churn.east] {
      $\forall_{ijk} a^{\farray}_{i} \pluseq B^{\farray\flist}_{ij} \cdot (d^{\farray}_{k} \cdot C^{\farray\flist}_{jk}),$ \\
      $\forall_{ijk} a^{\farray}_{i} \pluseq d^{\farray}_{k} \cdot (C^{\farray\flist}_{jk} \cdot B^{\farray\flist}_{ij}), \ldots$
    };
    \node [choice, below of=churn] (group) {enumerate where groupings (\ref{sec:schedule:group})};
    \node [example, right=of group.east] {
      $\forall_{ijk} \big(a^{\farray}_{i} \pluseq B^{\farray\flist}_{ij} \cdot w \where w \pluseq d^{\farray}_{k} \cdot C^{\farray\flist}_{jk}\big),$ \\
      $\forall_{ijk} a^{\farray}_{i} \pluseq d^{\farray}_{k} \cdot (C^{\farray\flist}_{jk} \cdot B^{\farray\flist}_{ij}), \ldots$
    };
    \node [choice, below of=group] (loop) {enumerate forall nestings (\ref{sec:schedule:nest})};
    \node [example, right=of loop.east] {
      $\forall_{j} \Big(\big(\forall_i a^{\farray}_{i} \pluseq B^{\farray\flist}_{ij} \cdot w\big) \where \big(\forall_k w \pluseq d^{\farray}_{k} \cdot C^{\farray\flist}_{jk}\big)\Big),$\\
      $\forall_{ijk} a^{\farray}_{i} \pluseq d^{\farray}_{k} \cdot (C^{\farray\flist}_{jk} \cdot B^{\farray\flist}_{ij}), \ldots$
    };
    \node [filter, below of=loop] (depth) {filter by max forall nesting depth (\ref{sec:schedule:depth})};
    \node [example, right=of depth.east] {
      $\forall_{j} \Big(\big(\forall_i a^{\farray}_{i} \pluseq B^{\farray\flist}_{ij} \cdot w\big) \where \big(\forall_k w \pluseq d^{\farray}_{k} \cdot C^{\farray\flist}_{jk}\big)\Big),$ \\ 
      $\big(\forall_{ij} a^{\farray}_{i} \pluseq B^{\farray\flist}_{ij} \cdot w\big) \where \big(\forall_{jk} w \pluseq d^{\farray}_{k} \cdot C^{\farray\flist}_{jk}\big), \ldots$
    };
    \node [pass, below of=depth] (workspace) {name workspaces and indices (\ref{sec:schedule:workspace})};
    \node [example, right=of workspace.east] {
      $\forall_{j} \Big(\big(\forall_i a^{\farray}_{i} \pluseq B^{\farray\flist}_{ij} \cdot w\big) \where \big(\forall_k w \pluseq d^{\farray}_{k} \cdot C^{\farray\flist}_{jk}\big)\Big),$ \\ 
      $\big(\forall_{ij} a^{\farray}_{i} \pluseq B^{\farray\flist}_{ij} \cdot w_j\big) \where \big(\forall_{jk} w_j \pluseq d^{\farray}_{k} \cdot C^{\farray\flist}_{jk}\big), \ldots$
    };
    \node [choice, below of=workspace] (protocolize) {enumerate protocols (\ref{sec:schedule:protocolize})};
    \node [example, right=of protocolize.east] {
      $\forall_{j} \Big(\big(\forall_i a^{\farray}_{\pinsert(i)} \pluseq B^{\farray\flist}_{\plocate(i)\plocate(j)} \cdot w\big) \where \big(\forall_k w \pluseq d^{\farray}_{\plocate(k)} \cdot C^{\farray\flist}_{\pstep(j)\plocate(k)}\big)\Big),$ \\ 
      $\big(\forall_{ij} a^{\farray}_{\pinsert(i)} \pluseq B^{\farray\flist}_{\plocate(i)\pstep(j)} \cdot w_{\pstep(j)}\big) \where \big(\forall_{jk} w_{\pinsert(j)} \pluseq d^{\farray}_{\plocate(k)} \cdot C^{\farray\flist}_{\plocate(j)\pstep(k)}\big), \ldots$
    };
    \node [filter, below of=protocolize] (filter) {filter by asymptotic complexity (\ref{sec:schedule:complexity})};
    \node [example, right=of filter.east] {
      $\big(\forall_{ij} a^{\farray}_{\pappend(i)} \pluseq B^{\farray\flist}_{\plocate(i)\pstep(j)} \cdot w_{\plocate(j)}\big) \where \big(\forall_{jk} w_{\pappend(j)} \pluseq d^{\farray}_{\plocate(k)} \cdot C^{\farray\flist}_{\plocate(j)\pstep(k)}\big),$ \\
      $\forall_{j} \Big(\big(\forall_i a^{\farray}_{\pinsert(i)} \pluseq B^{\farray\flist}_{\pstep(i)\pstep(j)} \cdot w\big) \where \big(\forall_k w \pluseq d^{\farray}_{\plocate(k)} \cdot C^{\farray\flist}_{\plocate(j)\pstep(k)}\big)\Big), \ldots$ 
    };
    \node [pass, below of=filter] (reformat) {add workspaces for transposes and reformatting (\ref{sec:schedule:reformat})};
    \node [example, right=of reformat.east] {
      $\big(\forall_{ij} a^{\farray}_{\pappend(i)} \pluseq B^{\farray\flist}_{\plocate(i)\pstep(j)} \cdot w^{\farray}_{\plocate(j)}\big) \where \big(\forall_{jk} w^{\farray}_{\pappend(j)} \pluseq d^{\farray}_{\plocate(k)} \cdot C^{\farray\flist}_{\plocate(j)\pstep(k)}\big),$\\
      $\forall_{j} \Big(\big(\forall_i a^{\farray}_{\pinsert(i)} \pluseq {B'}^{\farray\flist}_{\pstep(i)\pstep(j)} \cdot w\big) \where \big(\forall_k w \pluseq d^{\farray}_{\plocate(k)} \cdot C^{\farray\flist}_{\plocate(j)\pstep(k)}\big)\Big) \where \big(\forall_{ij} {B'}^{\flist\flist}_{ji} = B^{\farray\flist}_{ij}\big), \ldots$ 
    };
    \node [filter, below of=reformat] (autotune) {filter by empirical runtime (\ref{sec:schedule:autotune})};
    \node [example, right=of autotune.east] {
      $\big(\forall_{ij} a^{\farray}_{\pappend(i)} \pluseq B^{\farray\flist}_{\plocate(i)\pstep(j)} \cdot w^{\farray}_{\plocate(j)}\big) \where \big(\forall_{jk} w^{\farray}_{\pappend(j)} \pluseq d^{\farray}_{\plocate(k)} \cdot C^{\farray\flist}_{\plocate(j)\pstep(k)}\big)$\\
    };
    \node [program, below of=autotune] (exit) {scheduled program};
    \path [line] (init) -- (churn);
    \foreach \x in {-0.5,0,.5} {
      \path [line] ($(churn.south) + (\x cm, 0)$) -- ($(group.north) + (\x cm, 0)$);
    }
    \foreach \x in {-1,-0.5, 0, 0.5, 1} {
      \path [line] ($(group.south) + (\x cm, 0)$) -- ($(loop.north) + (\x cm, 0)$);
    }
    \foreach \x in {-1.5, -1.25, -1,-0.75, -0.5, -0.25, 0, 0.25, 0.5, 0.75, 1, 1.25, 1.5} {
      \path [line] ($(loop.south) + (\x cm, 0)$) -- ($(depth.north) + (\x cm, 0)$);
    }
    \foreach \x in {-1,-0.5, 0, 0.5, 1} {
      \path [line] ($(depth.south) + (\x cm, 0)$) -- ($(workspace.north) + (\x cm, 0)$);
      \path [line] ($(workspace.south) + (\x cm, 0)$) -- ($(protocolize.north) + (\x cm, 0)$);
    }
    \foreach \x in {-1.5, -1.25, -1,-0.75, -0.5, -0.25, 0, 0.25, 0.5, 0.75, 1, 1.25, 1.5} {
      \path [line] ($(protocolize.south) + (\x cm, 0)$) -- ($(filter.north) + (\x cm, 0)$);
    }
    \foreach \x in {-0.5,0,.5} {
      \path [line] ($(filter.south) + (\x cm, 0)$) -- ($(reformat.north) + (\x cm, 0)$);
    }
    \foreach \x in {-0.5,0,.5} {
      \path [line] ($(reformat.south) + (\x cm, 0)$) -- ($(autotune.north) + (\x cm, 0)$);
    }
    \path [line] (autotune) -- (exit);
\end{tikzpicture}
\caption{A visualization of the complete autotuning pipeline, where trapezoids
represent enumeration and filtering steps that add or remove programs from the
pipeline, and rectangles represent compiler passes that modify programs in the
pipeline. The corresponding subsections of Section \ref{sec:schedule} are
displayed in parentheses. Some examples of the result of each stage are shown on
the right. The arrows represent different programs at each stage, so more arrows
between stages represents a larger working set. It is important to filter as
early as possible to avoid large working sets. The last step is optional, and
could be expanded to several other steps which consider cache blocking or
parallelization, for example.} \label{fig:pipeline}

\end{figure*}

We use our asymptotic cost model to build an enumerative automatic scheduler.
Our scheduler works by making the most coarse-grained decisions first, such as
the structure of forall and where statements, working towards more fine-grained
decisions such as the formats and protocols of the tensors and workspaces. Our
scheduler stops once we have asymptotically optimized sequential programs, but a
more complete autotuner would make decisions regarding constant factors such as
register or cache blocking and/or parallelization. Because we enumerate all
possible choices at each stage of the pipeline, it is important to limit the
number of choices we make at each stage, and filter candidate programs between
stages so that the number of candidate programs does not grow too large. A
visualization of our approach is shown in Figure~\ref{fig:pipeline}.

\subsection{Enumerate Expression Rewrites}\label{sec:schedule:churn}
Our pipeline begins with a single pointwise index expression.  We then consider
associative and commutative rewriting transformations. These are rewriting rules
of the form $(a + b) + c \to a + (b + c)$ or $(a + b) \to (b + a)$. If the
expression contains $+$ and $\cdot$, we might also consider distributive properties
$(a + b) \cdot c \to a \cdot c + b \cdot c$ as well.  The purpose of all this rewriting is
to expose all grouping opportunities for the next stage.

\subsection{Enumerate Where Groupings}\label{sec:schedule:group}
Once we have a set of possible expression trees, we enumerate nontrivial ways that
where statements might be added to the expressions. These are transformations
of the form \[
  a_i \pluseq b_i \cdot c_i \cdot d_i \to a_i \pluseq b_i \cdot w \where w \pluseq c_i \cdot d_i,
\]
 where $+$ distributes over $\cdot$. We can also perform 
\[
  a_i \pluseq b_i \cdot c_i \cdot d_i \to a_i \pluseq b_i \cdot w \where w = c_i \cdot d_i,
\]
regardless of distributivity.  Note that we do not add workspaces for single
tensors, and only consider workspaces when there is an nontrivial operation
between more than one tensor whose result is to be recorded in the workspace.
When inserting workspaces at this stage, we do not yet need to consider the
indices that the workspace access needs, as this will be derived later from the
structure of foralls. We name the workspaces at this stage using De Bruijn
indexing. After inserting where statements, our naming scheme allows us to
normalize the pointwise expressions in each where and deduplicate our programs
somewhat.

\subsection{Enumerate Forall Nestings}\label{sec:schedule:nest}
We start by adding all the necessary foralls at the outermost level and then
moving these loops down into the pointwise expressions at the leaves of our
program. For instance, we might perform \begin{multline*}
\forall_{ijk} (a_{i} \pluseq B_{ij} + w \where w \pluseq C_{jk} + d_{k}) \to\\
\forall_{ij} ((a_{i} \pluseq B_{ij} + w) \where (\forall_k w \pluseq C_{jk} + d_{k})))\to\\
\forall_{j} ((\forall_i a_{i} \pluseq B_{ij} + w) \where (\forall_k w \pluseq C_{jk} + d_{k})))
\end{multline*}

When we move an index into a where statement, it may need to move into the
producer side, the consumer side, or both, depending on which side uses it and
whether there is a reduction operator in the producer expression. If the
producer has a reduction operator, then we move the loop into any side of the
where that uses that index in an access. Otherwise, we move the loop into the
consumer side always and into the producer side only if the producer uses that
index in an access.

After grouping the foralls, we can enumerate all permutations of contiguous
foralls.

\subsection{Filter by Maximum Nesting Depth}\label{sec:schedule:depth}

It is possible that our asymptotic cost model will not, for instance, recognize a triply-nested
loop as dominated by a double nested loop, because the triply-nested loop may
perform better on specific sparsity patterns. However, we believe that most
practitioners will want a program which performs well on relatively dense
uniformly random inputs, so we restrict our focus to programs with minimum
maximum nesting depth. We perform this filtering step as early in the pipeline
as possible to reduce the burden on subsequent stages. There has been extensive
research on heuristics to restrict the minimum maximum depth in tensor network
contraction orders, but our focus is somewhat different \cite{gray_hyper-optimized_2021, schindler_algorithms_2020, liang_fast_2021}. We are interested in
enumerating every program of minimum depth, and our programs are quite small.
This stands in contrast to prior work that focuses on finding a single minimum
depth schedule for a very large network. Thus, we chose our simple enumerative
approach.

\subsection{Name Workspaces and Indices}\label{sec:schedule:workspace}

At this point, we can give our workspaces fresh names and compute the indices we
need in their accesses. In the programs we have enumerated, a workspace linking
the producer and consumer sides of a where statement needs to be indexed by the
index variables shared by both the producer and consumer that are not quantified
at the top of the where statement. If we are scheduling for TACO, we can remove
workspaces with more than one dimension, since TACO does not support
multidimensional sparse workspaces and dense multidimensional workspaces would
be unacceptably large.

\subsection{Enumerate Protocols}\label{sec:schedule:protocolize}

At this stage, we normally enumerate all combinations of protocols for each mode
for each access. However, since TACO does not support hash formats, almost any
protocolization with more than two locates would require at least two
uncompressed level formats, resulting in an unacceptable densification of the
input.  Thus, when we schedule for TACO, we protocolize with all step or a
single locate at the first index to be quantified, the only two options that
would not densify inputs. If we wanted to consider densification, we might have
chosen to add the induced storage overhead to the asymptotic runtime, but we
reasoned that most practitioners would view these densifications as
unacceptable.

\subsection{Filter by Asymptotic Complexity}\label{sec:schedule:complexity}

Finally, we can use our asymptotic cost model to filter the protocolized 
programs. Note that we assume that tensors will eventually be permuted to match
the order in which index variables are quantified (concordant order). We also
ignore storage formats at this stage, as we will add those in the next stage, and
adding reformatting workspaces would needlessly complicate (but not change) the
asymptotic complexity.

\subsection{Add Workspaces for Transposes and Reformatting}\label{sec:schedule:reformat}

At this stage, we can reformat tensors and add formats to workspaces so that the
tensor access order is concordant (the level order of the sparsity tree matches
the order in which indices are quantified). Reformatting operations are achieved
by inserting workspaces that have the proper format. Since reformatting takes
linear time in the size of the tensor, we do not need to consider reformatting
in our asymptotic complexity. We choose formats for workspaces based on the set
of protocols they must support. For instance, a workspace that is written to via
append protocol and read via step protocol can be stored in uncompressed format,
but a workspace that is written with insert protocol and read via step protocol
must be stored in hash format. We also note that if there is a set of index
variables $i$ that are quantified at the top of the expression consuming a
tensor, all accesses to that tensor begin with $i$, and the corresponding modes
do not need reformatting, we can insert a workspace to reformat just the bottom
modes of the tensor. A similar sentiment holds for tensor writes. For example, consider
\[
\forall_{ikj}A^{\flist\flist}_{\pappend(i)\pinsert(j)} \pluseq B^{\flist\flist}_{\pstep(i)\pstep(k)} \cdot C^{\flist\flist}_{\pstep(k)\pstep(j)}.
\]
The last dimension of $A$ has an unsupported protocol, so we could reformat into
\begin{multline*}
  \big(\forall_{ij} A^{\flist\flist}_{\pappend(i)\pappend(j)} = w^{\flist\fhash}_{\pstep(i)\pstep(j)}\big) \where \\
  \big(\forall_{ikj} w^{\flist\fhash}_{\pappend(i)\pinsert(j)} \pluseq B^{\flist\flist}_{\pstep(i)\pstep(k)} \cdot C^{\flist\flist}_{\pstep(k)\pstep(j)}\big).
\end{multline*}

However, since the $i$ mode is quantified first at the top of the loop,
we can use a much simpler workspace

\begin{multline*}
  \forall_i \Big(
  \big(\forall_{j} A^{\flist\flist}_{\pappend(i)\pappend(j)} = w^{\fhash}_{\pstep(j)}\big) \where \\
  \big(\forall_{kj} w^{\fhash}_{\pinsert(j)} \pluseq B^{\flist\flist}_{\pstep(i)\pstep(k)} \cdot C^{\flist\flist}_{\pstep(k)\pstep(j)}\big)\Big).
\end{multline*}

Because TACO only supports dense one-dimensional \\ workspaces accessed with
step protocol, we compile outermost reformatting workspaces as explicit transposition and
reformatting calls, separate from the kernel. When compiling for TACO, we only
perform our workspace-simplifying optimization when it results in a
one-dimensional workspace.  TACO also only supports a single internal workspace,
so we filter kernels at this step with more than one workspace.

\subsection{Extensions and Empirical filtering}\label{sec:schedule:autotune}

At this point, one could employ additional cost models and transformations to
the asymptotically-good skeleton programs produced by the pipeline.
Transformations like parallelization, cache blocking, or register blocking might
be employed. One might imagine autotuning approaches that use input data to make
better informed choices between the remaining programs. Since these
transformations are out of scope, we simply run all programs in the frontier on uniformly
sparse square inputs and pick the best-performing one.

\section{Evaluation}

\begin{figure*}[htpb]
  \begin{tabular}{llrrrrl}
    Kernel &
    Description &
    \makecell{Min-Depth \\ Schedules} &
    \makecell{Undominated \\ Schedules} &
    \makecell{Min-Depth \\ Schedules \\ (TACO)} &
    \makecell{Undominated \\ Schedules \\ (TACO)} &
    \makecell{Asymptotic \\ Filter \\ Runtime} \\

    SpMV &
    $a_i = \sum B_{ij} \cdot c_{j}$ &
    \input{results/spmv_universe_length.json} &
    \input{results/spmv_frontier_length.json} &
    \input{results/spmv_tacoverse_length.json} &
    \input{results/spmv_tacotier_length.json} & 
    \input{results/spmv_tacoverse_mean_filter_time.json}\\

    SpMV\textsuperscript{2} &
    $a_i = \sum B_{ij} \cdot C_{jk} \cdot d_k$ &
    \input{results/spmv2_universe_length.json} &
    \input{results/spmv2_frontier_length.json} &
    \input{results/spmv2_tacoverse_length.json} &
    \input{results/spmv2_tacotier_length.json} & 
    \input{results/spmv2_tacoverse_mean_filter_time.json}\\

    SpMTTKRP &
    $A_{ij} = \sum B_{ikl} \cdot C_{kj} \cdot D_{lj}$ &
    3631104 &
    timed out
    &
    \input{results/smttkrp_tacoverse_length.json} &
    \input{results/smttkrp_tacotier_length.json} & 
    \input{results/smttkrp_tacoverse_mean_filter_time.json}\\

    SpGEMM &
    $A_{ij} = \sum B_{ik} \cdot C_{kj}$ &
    \input{results/spgemm_universe_length.json} &
    \input{results/spgemm_frontier_length.json} &
    \input{results/spgemm_tacoverse_length.json} &
    \input{results/spgemm_tacotier_length.json} & 
    \input{results/spgemm_tacoverse_mean_filter_time.json}\\

    SpGEMM\textsuperscript{2} &
    $A_{ij} = \sum B_{ik} \cdot C_{kl} \cdot D_{lj}$ &
    \input{results/spgemm2_universe_length.json} &
    \input{results/spgemm2_frontier_length.json} &
    \input{results/spgemm2_tacoverse_length.json} &
    \input{results/spgemm2_tacotier_length.json} & 
    \input{results/spgemm2_tacoverse_mean_filter_time.json}\\

    SpGEMMH &
    $A_{ij} = \sum B_{ik} \cdot C_{kj} \cdot D_{kj}$ &
    \input{results/spgemmh_universe_length.json} &
    \input{results/spgemmh_frontier_length.json} &
    \input{results/spgemmh_tacoverse_length.json} &
    \input{results/spgemmh_tacotier_length.json} & 
    \input{results/spgemmh_tacoverse_mean_filter_time.json}\\
  \end{tabular}
  \caption{Our test kernels, along with their descriptions and several
  statistics about our autotuning process. The ``Min-Depth Schedules'' column
  describes the number of schedules (with protocols and loop ordering for all
  tensors) of minimum maximum loop nesting depth. The ``Undominated Schedules''
  column describes the size of the frontier after asymptotic filtering of
  min-loop-depth schedules. The ``Min-Depth Schedules (TACO)'' and ``Undominated
  Schedules (TACO)'' columns are the same, but restrict schedules to what the
  TACO tensor compiler can generate. The asymptotic filter runtime is the
  average time (in seconds) to filter a single TACO-compatible min-loop-depth kernel.
  Filtering the full universe for SpMTTKRP timed out after a few days.}
  \label{fig:results_table}
\end{figure*}

We implemented our autoscheduler using the Julia programming language \cite{bezanson_julia_2012}. We
evaluate our approach using the TACO tensor compiler to run the generated
kernels. TACO does not implement protocolized concrete index notation in its
full generality, so we also ran a separate autoscheduling algorithm on the
subset of kernels supported by TACO.  After running a warmup sample to load
matrices into cache and jit-compile relevant Julia code, all timings are the
minimum of 10000 executions, or enough executions to exceed 5 seconds of sample
time, whichever happens first. We ran our experiments on an 12-core
Intel\textregistered Xeon\textregistered E5-2680 v3 running at 2.50GHz.
Turboboost was turned off. The generated TACO kernels were executed serially.

All inputs to kernels were in Compressed Sparse Fiber (CSF) format, meaning the
first mode was dense and all subsequent modes were sparse. Vectors were
therefore dense. All dimensions were the same size, and sparsity patterns were
uniform. We compared kernels in the asymptotic frontier using inputs of fixed
sparsity ($\rho=0.01$) with the first power-of-two dimension on which the
default kernel exceeded 0.1 seconds of runtime.

Figure \ref{fig:results_table} contains statistics describing the size of the
frontier before and after asymptotic filtering. As we can see, our asymptotic
cost model was usually able to reduce the cardinality of the universe of
min-loop-depth schedules by several orders of magnitude, even after restricting
to TACO-compatible schedules.  Our asymptotic filtering was able to process each
candidate schedule in less than half a second (this number accounts for
comparison between the candidate program and all other programs in the frontier,
as well as the time required to simplify the asymptote during preprocessing).
This time is independent of the size of the input to the program. To give a
sense of scale for the runtime, running a single sparse matrix-vector multiply
with TACO on a popular sparse matrix like ``Boeing/ct20stif'' of size $52,329
\times 52,329$ with $2,600,295$ nonzeros takes 9ms. When comparing so many
schedules using empirical runtimes, it's important to ensure that results are
statistically significant to avoid type 1 errors (false rejections of the null
hypothesis that kernels perform identically). As the number of kernels grows, so
too does the requisite number of samples required to compare them.  If we wanted
to run 100 trials on our single matrix, this would also take roughly a second
per schedule.  However, our method theoretically guarantees asymptotic
domination across all inputs, whereas empirical evaluation is specific to the
particular input matrix (or distribution of input matrices) under consideration.
Furthermore, we can evaluate asymptotic domination before making choices about
things like parallelization or cache blocking.

\begin{figure*}
  \begin{multicols}{2}
    \centering
    {\large SpMV $a_i = \sum B_{ij} \cdot c_{j}$}

    \hspace*{-0.05\linewidth}
    \begin{minipage}{1.1\linewidth}
    \includegraphics[width=0.5\linewidth]{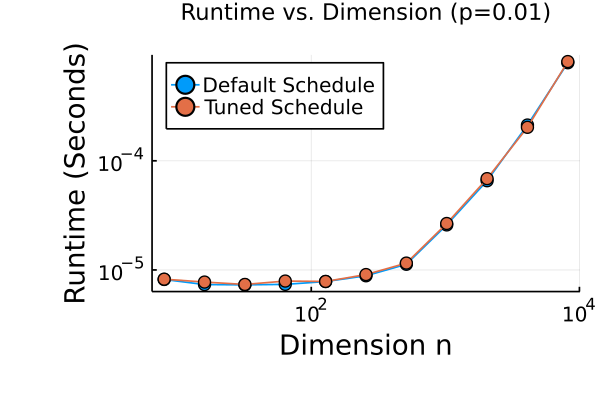}\includegraphics[width=0.5\linewidth]{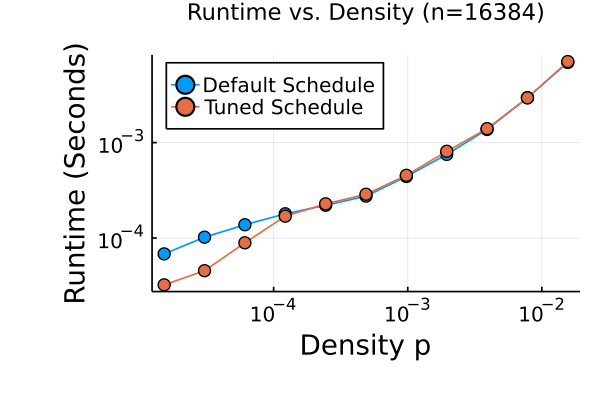}
    \end{minipage}

    {\large SpMV\textsuperscript{2} $a_i = \sum B_{ij} \cdot C_{jk} \cdot d_k$}

    \hspace*{-0.05\linewidth}
    \begin{minipage}{1.1\linewidth}
    \includegraphics[width=0.5\linewidth]{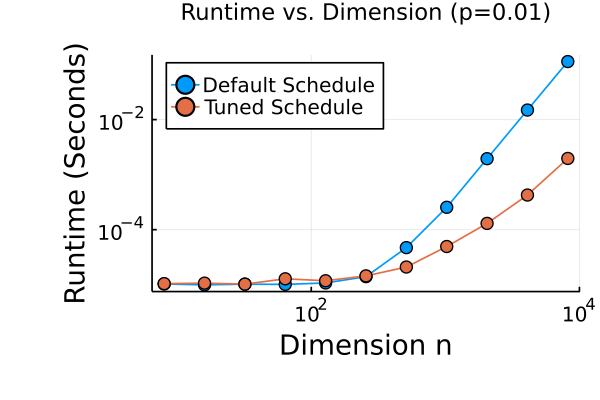}\includegraphics[width=0.5\linewidth]{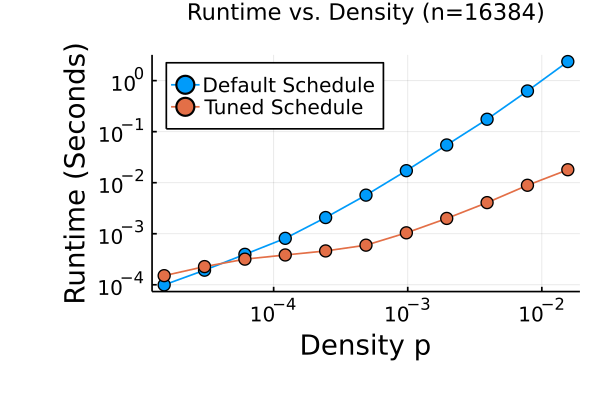}
    \end{minipage}

    {\large SpMTTKRP $A_{ij} = \sum B_{ikl} \cdot C_{kj} \cdot D_{lj}$}

    \hspace*{-0.05\linewidth}
    \begin{minipage}{1.1\linewidth}
    \includegraphics[width=0.5\linewidth]{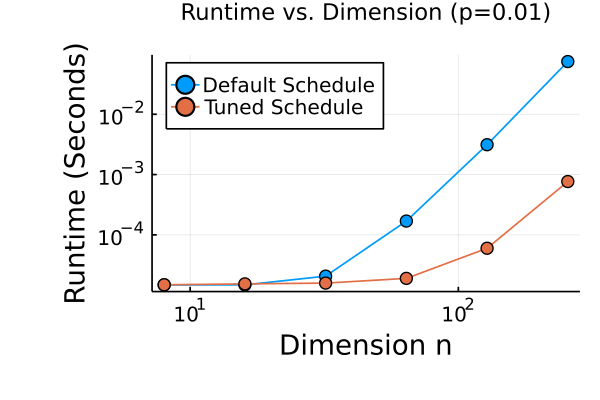}\includegraphics[width=0.5\linewidth]{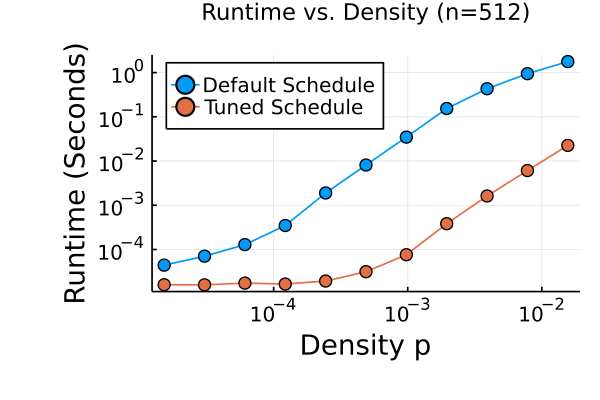}
    \end{minipage}
    \newpage
    {\large SpGEMM $A_{ij} = \sum B_{ik} \cdot C_{kj}$}

    \hspace*{-0.1\linewidth}
    \begin{minipage}{1.1\linewidth}
    \includegraphics[width=0.5\linewidth]{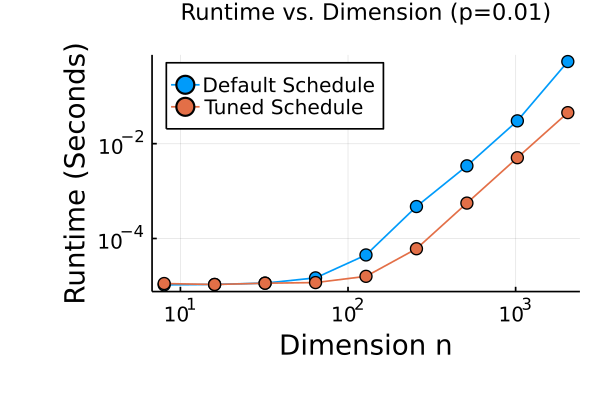}\includegraphics[width=0.5\linewidth]{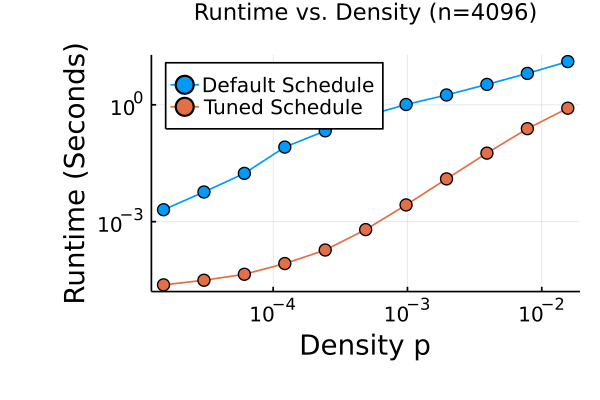}
    \end{minipage}

    {\large SpGEMM\textsuperscript{2} $A_{ij} = \sum B_{ik} \cdot C_{kl} \cdot D_{lj}$}

    \hspace*{-0.1\linewidth}
    \begin{minipage}{1.1\linewidth}
    \includegraphics[width=0.5\linewidth]{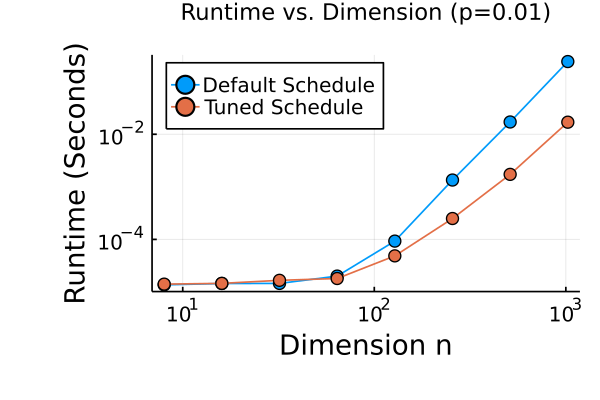}\includegraphics[width=0.5\linewidth]{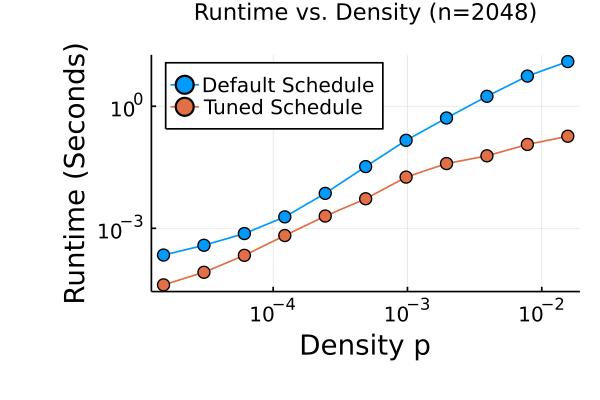}
    \end{minipage}

    {\large SpGEMMH $A_{ij} = \sum B_{ik} \cdot C_{kj} \cdot D_{kj}$}

    \hspace*{-0.1\linewidth}
    \begin{minipage}{1.1\linewidth}
    \includegraphics[width=0.5\linewidth]{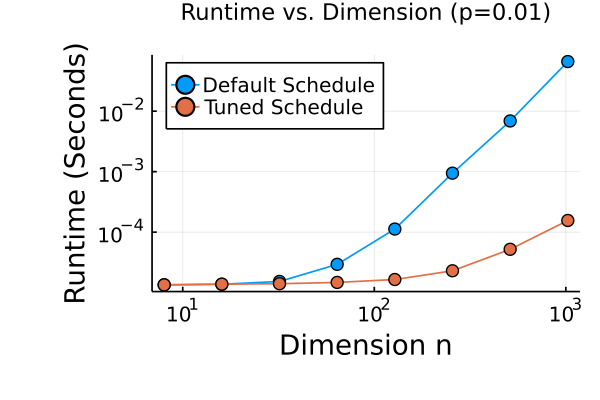}\includegraphics[width=0.5\linewidth]{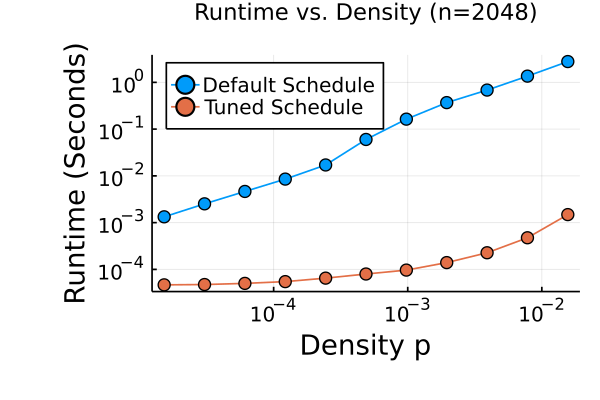}
    \end{minipage}
  \end{multicols}

  \caption{A comparison between the default schedule (nesting the loops in
  alphabetical order, without inserting any workspaces), and the schedule chosen
  by our autoscheduler on sparse inputs with uniformly random sparsity.}\label{fig:runtimecomparison}
\end{figure*}

Figure \ref{fig:runtimecomparison} compares
the autotuned schedules to the default schedules (created by nesting loops in
alphabetical order of the indices). Even in cases like SpGEMM or SpGEMMH, where
the default loop depth was at it's minimum (3 in this case), we see asymptotic
speedups due to improved reasoning about workspaces, loop ordering, and sparse
protocol accesses. Our autotuned kernels often improved on the defaults by
several orders of magnitude, and we saw increasing speedups as the dimension
increased. Interestingly, speedups sometimes increased and sometimes decreased 
with increasing density. In cases where speedups decreased with density (such as
SpMV\textsuperscript{2}), this may be due to the effects of sunk costs (like
initializing workspaces) beginning to become less noticeable as the compute
costs dominate. In cases where speedups increased with density, this
could be due to an autotuned kernel that takes advantage of improved filtering
opportunities, so that it would match the default kernel as inputs become dense.

\section{Related Work}

Decades of research have been devoted to the online autoscheduling problem for
sparse kernels, where optimizations are made to the kernel in response to the
sparsity pattern of the inputs. Notable examples include the OSKI sparse kernel
library \cite{vuduc_automatic_2004} and similar efforts to use analytical cost
models to choose between implementations, or more recent efforts to use machine
learning to make these decisions \cite{dhandhania_explaining_2021}. A broad
summary of similar efforts is given in \cite{strout_sparse_2018}.

Our efforts differ from such approaches as our decisions are made offline,
without examining matrix inputs. In the dense case, our problem looks quite
similar to the tensor contraction ordering problem, where the most important
cost function is simply the loop nesting depth and storage cost of dense
temporaries
\cite{gray_hyper-optimized_2021,schindler_algorithms_2020,liang_fast_2021}.
However, since we consider the sparsity of our tensors and kernels, our
algorithm is more similar to query optimizers for databases
\cite{chandra_optimal_1977,kolaitis_conjunctive-query_2000,konstantinidis_scalable_2013},
which use the theory of conjunctive query containment to reduce an input query
to its smallest equivalent.

Other related work combines sparse tensor algebra and relational
databases. Kotlyar et al. proposed implementing sparse tensor algebra using
relational algebra and database techniques
\cite{kotlyar_relational_1999,kotlyar_relational_1997-1,kotlyar_relational_1997,kotlyar_compiling_1997},
but did not give any techniques to optimize the generated code. It is unclear if
these techniques would extend to the additional formats and operations proposed
in \cite{chou_format_2018,henry_compilation_2021}.  Yu et al. proposed
integrating sparse matrix operations into relational queries, and proposed a few
techniques to optimize the resulting join structures. However, this technique
was not generalized beyond matrices, nor does it apply directly to tensor
compilers and the kinds of programs they generate\cite{yu_scalable_2021}.

Set-like representations have been used before to describe implementations of
tensor programs. The most popular example of this is the polyhedral framework,
which represents the iterations of dense loops with affine loop bounds as
abstract geometric polyhedra. We leave a summary of such prior work to
\cite{baghdadi_tiramisu_2019}. The polyhedral model has been extended to the
sparse case by Strout et al.
\cite{strout_approach_2016,strout_sparse_2018-1,rift_optimizing_2021-1}.
However, the sparse polyhedral representations are used to describe the schedule
itself, and not the complexity of the schedule. These sparse polyhedral
representations have therefore not been used to automatically analyze or
optimize the asymptotic complexity of program implementations.

\section{Conclusions}

Our cost model represents the first automatically derived expression of
asymptotic complexity for sparse tensor programs. We describe algorithms to
determine when one program asymptotically dominates another. Our enumerative autoscheduler
can be used to produce a frontier of asymptotically sound skeleton programs as a starting
point for further optimizations, such as parallelization or cache blocking.

Our cost model can be extended in several ways. By emitting set expressions for the
tensor patterns, we can analyze the intermediate storage size of various tensor
temporaries. By stratifying our tasks into multiple sets (e.g. a set for
expensive tasks and a set for cheaper ones), we can begin to abstractly analyze
some of the constant or logarithmic performance differences between, say,
floating point operations and hash table operations. Our model can
also be applied to programs with nonzero fill values or new semirings.

Our cost model also has applications beyond an enumerative autoscheduler. If
better heuristics are to be developed for optimizing sparse tensor programs, our
model can be used to evaluate them. Our model might also be used to guide a more
sophisticated schedule search strategy, like greedy approaches, simulated
annealing, or learning-based approaches. In the future, our offline techniques could be used to enhance online techniques,
using runtime data to choose among the templates produced by the autoscheduler.

Sparse tensor compilers cannot expect end users to schedule their own programs. Automatic scheduling is necessary to fully abstract the details of sparse tensor compilation. The offline asymptotic decision making described in this paper is the last piece of infrastructure needed to make sparse tensor compilation accessible to the mainstream. 

\begin{acks}                            
  This work was supported by a Department of Energy Computational Science
  Graduate Fellowship, DE-FG02-97ER25308.
\end{acks}

\bibliography{Asymptuner}

\end{document}